# Using Deep Neural Network for Android Malware Detection

Abdelmonim Naway[1], Yuancheng LI[1]
[1]North China Electric Power University, School of Control and Computer Engineering, 2 Beinong Road, Chanping District, Beijing, China,102206
{abdelmonim, yuanchengli}@ncepu.edu.cn

*Abstract*—The pervasiveness of the Android operating system, with the availability of applications almost for everything, is readily accessible in the official Google play store or a dozen alternative third-party markets. Additionally, the vital role of smartphones in modern life leads to store significant information on devices, not only personal information but also corporate information, which attract malware developers to develop applications that can infiltrate user's devices to steal information and perform harmful tasks. This accompanied with the limitation of currently defenses techniques such as ineffective screening in Google play store, weak or no screening in third-party markets. Antiviruses software that still relies on a signature-based database that is effective only in identifying known malware. To contrive with malicious applications that are increased in volume and sophistication, we propose an Android malware detection system that applies deep learning technique to face the threats of Android malware. Extensive experiments on a real-world dataset contain benign and malicious applications uncovered that the proposed system reaches an accuracy of 95.31%.

## I. INTRODUCTION

Nowadays, noticeable amid the most substantial devices you can use to assist with your life is a smartphone. There are numerous things that you can accomplish with a smartphone in various ways (checking email, messaging companions, checking the climate and news, making appointments, making bank transactions, etc.). Smartphone corporations dispatched a sum of 1.46 billion devices in 2017. The Android seized around 85% of the overall smartphone volume [34]. According to IDC, five out of six new phones were running Android. Ericsson forecasts there could be upwards of 6.4 billion smartphone users close to the ending of 2020, nearly one for each individual [10]. Smartphones are a progressively appealing focus for online criminals. Therefore, they are putting resources into more advanced assaults that are powerful at taking precious individual information or blackmailing money from prey. Android users persist as the dominant victim of malicious developers. SophosLabs has processed the approximately 10 million Android samples submitted by Sophos clients for investigation before the end of 2017. This is up from the 8.5 million handled through all of 2016. For 2017, the number was up to almost 3.5 million. By far most are really malicious, with 77% of the submitted instances ending up being malware [15]. At the beginning of 2018, the G DATA security professionals recognized an average of 9,411 advanced malware daily for the Android. This implies a new malware emerging every 10 seconds. For the entire 2018, the G DATA analysts are estimating around 3.4 million new Android malware [22]. The situation even worsens by the fact that some adware is pre-installed on a few hundred various Android-powered device models and versions, according to the Avast Threat Labs [44].

The bulk of accessible applications in the Google Play Store were most as of late set at 3.3 million applications in March 2018 [4]. In May 2017, Google reported its new implicit malware defense for Android, Play Protect, which verifies applications and APK files whenever they are downloaded utilizing the Google official store or third-party stores. Since August 2017 and afterward, it has been accessible on all Android devices with Google Play Services 11 or above, and is set up on devices with Android 8.0 and above. However, when Play Protect is tested, it's only able to detect 51.8% of the test cases [37]. Another defense mechanism against Android malware is Android Antivirus. The principle disadvantage of antivirus applications is that they depend on the signature-based database for malware detection [18]. While signature-based antivirus is efficient in identifying known malware, they come short in distinguishing new malware. Besides they can be overcome by changing the manners by which an attack is made. To overcome signature-based limitations, heuristic scanning is proposed, which uses rules to search for commands that may suggest malicious intentions. But malware can evade heuristic scanning if they manage to mask their malicious action [47]. To settle the problem of heuristic search, machine learning methods are proposed. Machine learning approaches are able to find certain patterns to identify formerly not observed malware instances. The increasing complication of Android malware requires a new detection strategy that is robust enough to be circumvented. To this end deep learning as a subset of machine learning introduced to detect Android malware.

In this work, we inspect the use of the deep neural network to Android malware detection by extracting features from AndroidManifest file and java files that fed into the deep neural network (DNN) classifier for classification of apps into benign or malicious. Real-world samples have been collected from different open sources. A comprehensive test is performed to evaluate the effectiveness of the proposed system. Results unveil that the schemed system is competent to identify malware. The major contribution of this study to the development of malware detection techniques using deep learning are as follows:
- The proposition of a novel Android malware detection approach builds on deep learning
- implementation of a model that performs static analysis to provide five different feature sets: permission combinations; API calls; intent filters; valid certificates; the presence of APK files in the asset folder. To the best of our knowledge, this is the first time these features are synthesized for Android malware detection
- Collection of malware dataset that involves various malware types from 2013 up to 2017 and benign dataset the contain different application categories
- A comprehensive test of the system that includes evaluating the system with different features set, comparing the performance of the DNN with commonly machine learning methods, and comparing the system with some other works in the literature.

The rest of the paper is organized as follows: Section 2 relates the concerned works done Android malware detection. The proposed methodology is demonstrated in Section 3. Followed by Section 4 which shows and discusses the results of comprehensive experimenting. Finally, Section 5 concludes the study by underlining important points.

## II. RELATED WORK

Lately, the utilization of machine learning and deep learning methods into the analysis of Android malware has grown into a dynamic research field. This section reviews some of the presently concerned works of machine learning and deep learning for static analysis and dynamic analysis in Android malware detection and classification.

1. Android Malware Detection using Deep learning

Zhenlong Yuan et al. in [43], explored the use of deep learning in Android malware identification by using static and dynamic features extracted from APK files. The authors developed a deep belief network (DBN) model for classification utilizing small dataset and achieved high detection accuracy. Zhenlong Yuan et al. [31], developed Droiddetector based on their previous work [43], an Android detection system based on deep learning that can identify Android malware automatically by using large dataset and numerous features. The results showed that deep learning methods can accomplish remarkable accuracy, surpassing conventional machine learning methods. Lifan Xu et al. [35], developed HADM a hybrid Android malware classification method. The deep autoencoder was used to study static and dynamic apps features. Then support vector machines utilized for classification. The performance of the model was evaluated separately for static features and dynamic features. Static features provide much better accuracy than dynamic features.

Shifu Hou et al. [40], designed DroidDelver an automatic Android malware characterization system employing DBN depending only on extracted API call blocks from the smali code. Testing results illustrated the high performance of DBN over shallow architecture methods. Xin Su et al [19], proposed DroidDeep a malware identification scheme for the Android platform relying on static features. The DBN was used to study the features and Support Vector machine implemented for malware classification. The results demonstrated high performance not merely in the test dataset but also in comparison with similar approaches. Shifu Hou et al. [3], developed an automated Android malware detector Deep4MalDroid. Resting on a graph representation of extracted Linux system call, the Stacked Autoencoders (SAEs) was used to study common patterns of malware and hence to identify newly unknown malicious applications.

Niall McLaughlin et al. [8], motivated by the success of a convolutional neural network (CNN) in natural language processing, presented an Android malware detection system based on CNN. They employed opcode sequences to perform malware classification. A series of experiments were conducted, and the results demonstrated good accuracy in the testing dataset but, the accuracy dropped when the system tested on a new dataset larger than the testing dataset. Ton et al. [48] developed a color- heightened CNN for malware detection. They built color representations for transforming Android apps into RGB color code and interpret them into a defined sized encoded image. Then the encoded image was forwarded to CNN for automated feature drawing and learning. Although the main objective was to create an automated system, results suggest that sample collection and model updates are still required. R. Vinayakumar et al. [45], examined the use of long short-term memory (LSTM) for Android malware identification. They performed several tests using various LSTM (a technique to check long-range dependencies through time-varying patterns) network topologies with several network parameters on extracted features employing static and dynamic analysis. The reported results indicated a good performance of the LSTM in comparison with the recurrent neural network and shallow machine learning architecture.

2. Android Malware Detection using Machine Learning

Andrew Bedford et al. [51] developed ANDRANA a malware detection approach based on machine learning algorithms. ANDRANA utilizes static analysis to identify features, and Machine Learning algorithms to distinguish if these features are good enough to decide an application as a malware. ANDRAN analyzes applications in three stages. In the first phase, the app was disassembled to acquire its code. Then, applying static analysis, the application's features are drawn. Finally, a classification algorithm determines from the set of features if the application is malicious. Borja Sanz et al. [52] proposed Manifest Analysis for Malware detection in Android (MAMA). A method for Android malware detection. This approach was based on certain features obtained from the Manifest file of Android applications. Specifically, they utilized the permissions and the feature tags within the manifest file. These features are then employed to construct supervised machine-learning algorithms to identify malicious applications.

A. Feizollah et al. [27] designed AndroDialysis a malware identifier depending on the analysis of two different sorts of Intents (implicit and explicit Intents) extracted from benign and malicious apps. The performance of the system was evaluated using Intent only which achieved 91% accuracy versus 83% using permissions. Combination of Intent and permission led to better detection accuracy. Babu et al. [53] proposed DROIDSWAN a system for Android malware detection, established on static analysis of Android APK. DroidSwan extracts numerous essential features from an Android application gives weight to these features and constructs a classification model utilizing machine learning algorithms. S. Y. Yerima et al. [25] developed a malware detection system based on various features to build multiple machine learning classifiers in parallel with the purpose of obtaining better accuracy. Hui Juan et al. [54] proposed HEMD Android malware characterization system built on Random Forest and utilizing diverse static features. The best accuracy obtained by HEMD was 89.91%.

Mas' ud et al. [55] designed malware detection system using dynamic analysis resting on five different set of features and applying five different machine learning classifiers with the objective of acquiring the best combination for efficient Android malware identification. The experimental results show that a multilayer perceptron classifier provides the best accuracy.

## III. THE PROPOSED METHODOLOGY

### 1. Android Application Overview

Android applications (also refers to as apps) are principally written in Java. The Android SDK set of development and debugging tools for Android arrange the code alongside resource files and data into an APK, an Android package file format with an APK suffix. One APK file includes all of the constituents of an Android app and is the file that Android-controlled devices use to set up the application [11]. There are four fundamental components namely Activities; services; broadcast receivers; content providers that formulate the Android application. Every component represents an entry point for which the system or a client can enter the application. Prior to run any of the application components, the Android system must recognize that the component occurs by parsing the application's AndroidManifest.xml file. It is a precondition for the application to specify each component in this file, and obliged to be at the base of the app project directory.

Reverse engineering is a technique exploited to analyze Android malware. In this procedure, the application is decompiled to understand its working and functionality by analyzing the code. When APK file decompiled, it contains the following components [5], [6]:

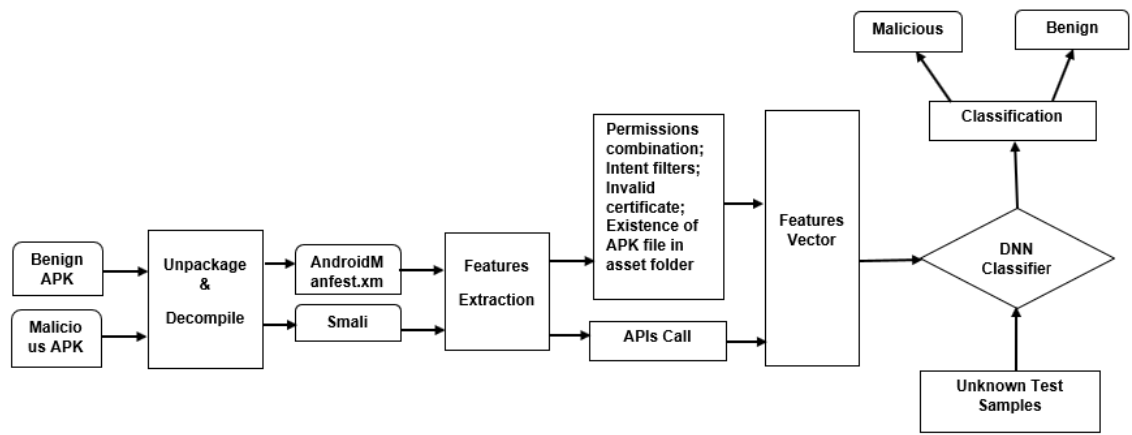

Figure 1. Illustrate the proposed methodology

- Meta-inf Folder: comprise of data that enables clients to ensure the security of the Android and completeness of the APK application.
- Res Folder: comprise XML specifying the layout, attributes, languages, etc...
- Classes.dex: hold the whole Java source code that is compiled. This file is run on the Android Runtime (ART). This file comprises the entire bytecode that ART will compile.
- AndroidMainifest.xml File.
- Resources.arsc: this file is a binary resource file that is obtained after compilation

### 2. The Structure of The Proposed Methodology

Fig. 1 outline the structural design of the proposed methodology. The proposed methodology consists of the following steps:
- Apps de-compilation: The Mobile Security Framework (MobSF) [16] which, is a powerful tool for Android malware analysis that is capable of retrieving the code from obfuscated apps was used to unpackage and decompile the APK files

- Features Extraction: MobSF is also used to extract features from two files, first, Androidmanifest.xml, which include the Permissions and Intent filters. Then the APIs calls are drawn from smali files generated by MobSF. The asset folder is screened for the presence of APK files. Finally, the legitimacy of the certificate is verified
- Feature Vector: All of the extracted features are transformed into a feature vector.
- DNN Classifier: Building upon the extracted features the DNN study features to characterize Android apps
- Classification: By applying the DNN classifier, the new unknown apps will be labeled as benign or malicious.

3. Features Set

**Permissions combination(fs1)**: permission is confinement restricting access to a portion of the code or data on a device. The restriction is enforced access to defend vital data and code that could be abused to harm the user's experience. Permissions are employed to grant or constrain an application access to restricted APIs and resources [46]. Permissions that are requested by an application can suggest its functionality to some extent. However, permissions can service both malicious apps and benign ones. In this manner, it is essential to combine various permission requests together for more accurate malware identification [20], [29].

**Intent Filters(fs2)**: The intent is an informing object that can be implemented to demand an action from another app component and is regarded as a security method to block applications from attaining access to other applications directly. Malware applications frequently listen to particular intents [7], [27].

**API Calls(fs3)**: APIs used by an application decides the real usefulness and capacity of the application. Static examines of the APIs handled in an application is crucial to empathize with what the application really plans to do. Specific API calls allow access to system services or resources of the device and are often discovered in malicious applications. As these calls can be particularly direct to malicious behavior [28]. The following API calls are examined: Telephony Manager. [25]; HTTP connection and sockets [36]; DexcClassLoader [32]; reflection; System Service; Runtime and System; cryptographic operations (crypto) [23].

**Invalid certificate(fs4)**: To verify the validity of the certificate. An invalid certificate proves that the application has been modified(repackaged) [12].

**Presence of APK (fs5)** files in the asset folder: Concealing data in asset folder is considered a typical behavior of the malicious application. [38]

4. Feature Vector

All of the extracted features are transformed into a feature vector. Each feature set is defined as a Boolean expression with different dimensions and then generates a unified representation by mapping different sets into a joint vector space. For evaluation, the defined combination of defined feature set S:

$$S = fs1 \cup fs2 \cup fs3 \cup fs4 \cup fs5 \quad (1)$$

$|Si|$- dimension set Si is used to represent a feature set which is a vector of zeros with a '1' in the position if malware has used a certain feature in the feature set. Therefore, we can convert any application X to a vector space $\varnothing(x)$:

$$\varnothing : x \rightarrow \{0,1\}^{|s|}, \varnothing(x) \rightarrow (I(x,s))s \in S \quad (2)$$

Where the indicator function $I(x,s)$ is defined as:

$$I(x,s) = \begin{cases} 1 & \text{if the application } x \text{ contains feature } s \\ 0 & \text{otherwise} \end{cases}$$

Therefore, different features can be mapped to into unified joint vector.

5. Deep Neural Network (DNN)

Deep learning lets computational models that are compiled from different processing layers to find out representations of data with different levels of abstraction. Deep learning recognizes sophisticated structure in substantial data sets by applying the back-propagation algorithm to demonstrate how a machine should adjust its inside parameters that are applied to calculate the representation in every layer from the representation in the former layer [30]. Deep neural networks are deep learning methods, which enable them to study complicated nonlinear functions of a given input for the purpose of decreasing error value.

The DNN used to implement a classification task may have the following generic architecture: an input layer, which is provided with some input vectors constituting the data; 2 or more hidden layers, where a transformation is utilized to the output of the former layer, receiving a higher-level representation as we go off from the input layer; and an output layer, which calculates the output of the deep neural network. In this final layer, the output is equated (with regard to supervised learning) to the class label (true value) and the error standard is employed to calculate the cost [39].

The model is determined by its arguments: weight matrices $W_{j, j-1}$, and bias vectors $b_j$, with j moving from 1 to the number of hidden layers. These arguments are tuned repeatedly to minimize the cost function, commonly with stochastic gradient descent. Then, an assigned training set $(x^{(i)}, y^{(i)})$, where $x^{(i)}$ is a stated feature vector and $y^{(i)}$ its comparable class, each hidden layer implements a non-linear translation function g to the output of the former layer. This translation considers the arguments W and b which associate one layer to its former one, and renders the activation values of neurons with the next equations [39]:

$$h_j(x^i) = g(w_{j,j-1} h_{j-1}(x^{(i)} + b_j)), j = 2,...,N-1 \quad (3)$$

$$h_1(x^i) = g(w_{0,1} h_{j-1} x^{(i)} + b_1) \quad (4)$$

Lastly, for a classification job, the output layer calculates a softmax function, which produces the probability P of a given input x to pertain to a certain class c:

$$P(c \mid h(x)) = \frac{\exp(w_l^c h_l(x) + b_l^c)}{\sum_{k=1}^{c} \exp(W_l^k h_l(x) + b_l^k)} \quad (5)$$

where hl(x) indicates the final hidden layer activation for input x, and stand for the weights matrix and bias vector recipient, which link the output unit for class c with the final hidden layer, and C is the whole number of classes. To tune the arguments to the job, a cost function is regarded, attempting to minimize the error between the prediction (network output) and the true class, and arguments are adjusted gradually through backpropagation [39].

### 6. Dataset Setup

For all the experiments, we look at a dataset of real Android apps and malware. The dataset contains 1200 apps in total, including 600 benign apps and 600 malicious apps. 570 apps out of the 600 benign apps were downloaded from the official Google play store in September 2017. The rest of the apps were downloaded from a third market APKpure [49], within the same period. The benign apps belong to different categories, as described in Table 1. And exemplify popular apps in their respective category (All the downloaded apps have at least 500,000 installations). All benign apps passed the scan of the latest versions of 3 Antivirus, ESET Internet Security 11.2, Symantec Endpoint 14, and Kaspersky Internet Security 18.

The malware samples come from different open sources as detailed in Table 2. Malicious apps represent different malware types such as Trojan horses, Backdoors, Information Stealers, Ransomware, Scareware, Fakeware, and remote access Trojans. The malware was selected from samples collected from May 2013 up to December 2017.

## IV. RESULT AND DISCUSSION

After a thorough demonstration of the proposed system, we now go forward to an empirical evaluation considering the performance of the proposed system.

### 1. Experiments Set Up

The dataset is split into a training set (80%) and a validation set (20%), and the deep neural network model is built using Keras utilizing TensorFlow's backend. Fig. 2 describes the structure of the deep neural network. Relu and softmax are used as activation functions. Gradient Descent is used for optimization. Mean Squared Error is used to measure the error.

### 2. Malware Detection

In this experiment, the proposed system malware detection is evaluated according to the performance metrics described in Table 3. The results are shown in Table 4.

```
Layer (type)                 Output Shape              Param #
=================================================================
dense_1 (Dense)              (None, 250)               10250
dense_2 (Dense)              (None, 200)               50200
dense_3 (Dense)              (None, 150)               30150
dense_4 (Dense)              (None, 100)               15100
dense_5 (Dense)              (None, 2)                 202
=================================================================
Total params: 105,902
Trainable params: 105,902
Non-trainable params: 0
```

Figure 2. Illustrate DNN parameters

TABLE 1. BENIGN APPS CATEGORIES

| Apps categories and subcategories | Number of Apps |
|---|---|
| Security (Antivirus- Password managers- VPN- Parental lock- Lock screen) | 60 |
| Books and magazines | 24 |
| Web browsers | 21 |
| Business and finance | 23 |
| Education | 32 |
| Family (parenting) | 12 |
| Food | 14 |
| Games (Strategy- Race- Music- Board -Sport-Casino) | 53 |
| Health and fitness | 23 |
| Image editing | 42 |
| Launchers | 19 |
| Maps | 20 |
| Medical | 16 |
| Messengers | 23 |
| Multimedia | 26 |
| News | 23 |
| Office | 24 |
| Music and podcast | 21 |
| Shopping | 19 |
| Sport | 20 |
| Translations | 10 |
| Utilities | 59 |
| Videos | 16 |
| Total | 600 |

TABLE 2. MALWARE SAMPLES SOURCES

| Source | NO. of samples |
|---|---|
| Contagio Mobile malware minidump [42] | 44 |
| DroidBench [17] | 30 |
| GitHub- Android malware master [14] | 80 |
| Viruschare [13] | 300 |
| Virussign [50] | 146 |
| Total | 600 |

TABLE 3. PERFORMANCE METRICS

| Terms | Description |
|---|---|
| True Positive (TP) | Number of samples correctly distinguished as malicious |
| True Negative (TN) | Number of samples correctly distinguished as benign |
| False Positive (FP) | Number of samples erroneously distinguished as malicious |
| False Negative (FN) | Number of samples erroneously distinguished as benign |
| Accuracy | (TP+TN)/(TP+TN+FP+FN) |
| Precision | TP/(TP+FP) |
| Recall | TP/(TP+FN) |
| F1-Score | 2*((precision*recall)/ (precision +recall)) |

TABLE 4. THE PERFORMANCE RESULTS OF THE DNN

| Accuracy | Precision | Recall | F1-Score |
|---|---|---|---|
| **95.31** | **95.35** | **95.31** | **95.31** |

Table 4. Demonstrate that the ability of the proposed system to identify an app as benign or malicious is 95.31%. The precision of correctly classified apps within all classified apps is 95.35%. The ratio (recall) of correctly identified benign samples is 95.31%. The F1-Score which points to how much the model determinate is 95.31%.

Fig. 3 show the loss in training and validation sets which is remained small and reflect the good performance of the model.

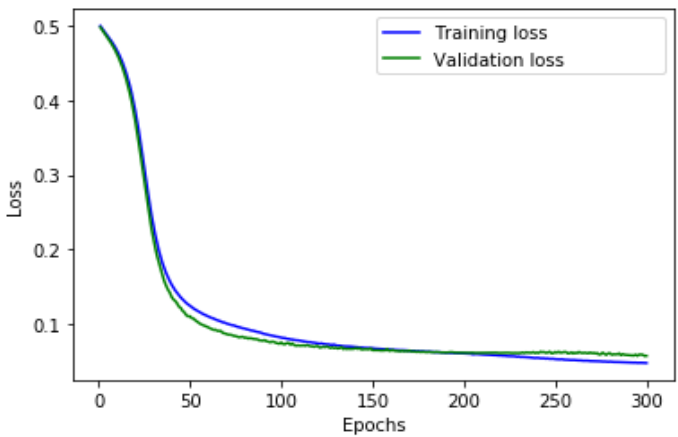

Figure 3. The loss in training and validation sets

Furthermore, Fig. 4 display the accuracy in training and validation sets that indicate the good performance of the model as well.

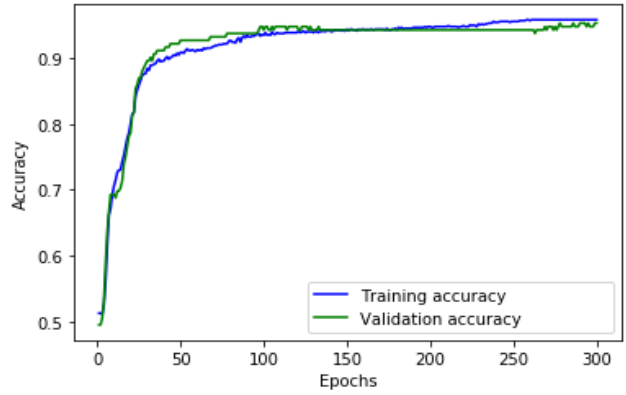

Figure 4. Accuracy in training and validation sets

3. Comparisons of Detection Using Different Features Set

In this experiment, the performance of malware identification is compared among the features set. The results are interpreted in Table 5.

TABLE 5 EVALUATION OF THE DNN USING DIFFERENT FEATURES SET

| Features Set | Accuracy |
|---|---|
| All Features Set | **95.31** |
| APIs Only | 90.10 |
| Permissions Combination Only | 82.29 |
| Intent-filters Only | 69.79 |
| APIs + Permissions combination + Invalid Certificate + Presence of APK file in asset folder | 93.22 |
| APIs + Intent-filters + Invalid Certificate + Presence of APK file in asset folder | 91.14 |
| Permissions Combination+ Invalid Certificate + Presence of APK file in asset folder | 83.33 |

From the results in Table 5. We can note that the best accuracy is achieved when all features are used in the classification process. Using APIs + Permissions combination + Invalid Certificate + Presence of the APK file in the asset folder come second best with 93.22% while, using APIS only yielded 90.10%. Permissions combination alone gives only 82.29%, adding the Invalid Certificate + Presence of the APK file in the asset folder that raised its accuracy to 83.33%. Using Intent Filters only provides the poorest accuracy among all features set. This is because there are many Intent filters used by both malicious and benign apps. Thus, the accuracy increases when Intent filters are used in combination with some other features.

4. Comparisons of Common Machine Learning Methods in Malware Detection

In this experiment, the performance of the proposed system is compared with 4 common machine learning methods Decision Tree (DT), K-nearest neighbor (KNN), Random Forest (RF), and Support vector machines (SVM) using the same setting applied to evaluate DNN. The results are displayed in Fig. 5.

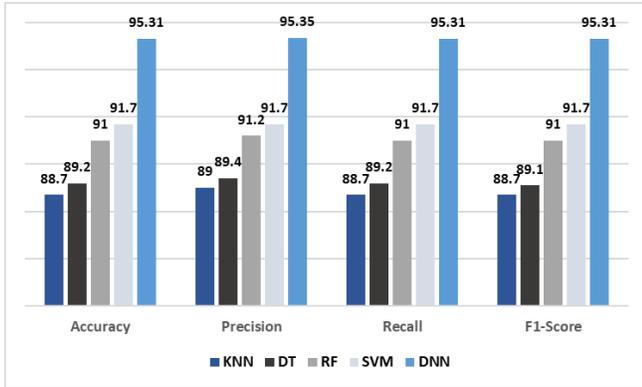

Figure 5. Comparison of common machine learning method with the proposed system

Fig. 5 shows that the proposed system exceeds the common machine learning methods in all performance metrics. SVM attained the second-best accuracy with 91.7%. The KNN produced the least accuracy 88.7% among all classifiers.

5. Comparison with Some Other Works in Literature

In this experiment, further, the proposed system is compared with some other works in literature in terms of accuracy. The result of the comparison is explained in Table 6.

TABLE 6. COMPARISON OF THE PROPOSED SYSTEM WITH SOME OTHER WORK IN THE LITERATURE

| *Approach* | *Accuracy* |
|---|---|
| This work | 95.31 |
| W. Li et al. [33] | 90 |
| M. Ganesh Li et al. [9] | 93 |
| H. Liang et al. [24] | 93.16 |
| F. Martinelli et al. [2] | 90 |
| H. Alshahrani et al. [1] | 95 |
| L. Xu et al. [35] | 94.7 |
| S. Hou et al. [26] | 96.66 |
| S. Hou et al [3] | 93.68 |
| T. Huang et al. [48] | 93 |
| Z. Wang et al. [21] | 93.09 |

As can be noted from Table 6. The proposed system provides better accuracy than the approaches in the literature except [26], which, has a slightly higher accuracy than the proposed system.

V. CONCLUSION

The unprecedented increase of Android malware in a previous couple of years, besides the drawbacks in current protection mechanism, requires an efficient solution to restrain Android malware. In this work, we propose a system for Android malware detection using the deep neural network, which uses permissions combination, Intent filters, Invalid certificate, the existence of APK file in asset folder, and API calls as features to construct a deep neural network (DNN) that is capable of identifying malicious from benign ones. Comprehensive test results with different features set reveal that the proposed system achieves 95.31% accuracy. The performance of the system is compared with common machine learning methods, results show the proposed system surpasses the machine learning methods in all performance metrics. Finally, the performance of the proposed system is equated with some other work in literature, in terms of accuracy, the results denote that the proposed system delivers better accuracy than most recent work.

As future work, we look for more improvement in the accuracy of malware detection by exploring other deep learning methods.